\documentclass[prl,twocolumn,showpacs,superscriptaddress]{revtex4-1}
\usepackage{graphicx,amssymb,amsmath,color,psfrag,bbm}
\begin{document}

%\definecolor{darkgreen}{rgb}{0,0.5,0}

\title{Quadratic band touching with long range interactions in and out of equilibrium}
\author{Bal\'azs D\'ora}
\email{dora@eik.bme.hu}
\affiliation{Department of Physics and BME-MTA Exotic  Quantum  Phases Research Group, Budapest University of Technology and
  Economics, Budafoki \'ut 8, 1111 Budapest, Hungary}
\author{Igor F. Herbut}
\affiliation{Department of Physics, Simon Fraser University, Burnaby, V5A 1S6, British Columbia, Canada}

\date{\today}

\begin{abstract}
Motivated by recent advances in cold atomic systems, we study the equilibrium and quench properties of two dimensional fermions with
quadratic band touching at the Fermi level, in the presence of infinitely long range interactions. Unlike when only short range interactions are present,
both nematic and quantum anomalous Hall (QAH) states state appear at weak interactions, separated by a
narrow coexistence region, whose boundaries mark second and third order quantum phase transitions.
After an interaction quench, the QAH order exhibits three distinct regions: persistent or damped
oscillations and exponential decay to zero. In contrast, the nematic order \emph{always} reaches a non-zero stationary value through power
law damped oscillations, due to the interplay of the symmetry of the interaction and the specific topology of the quadratic
band touching.
\end{abstract}


\maketitle

\paragraph{Introduction.}
Quantum quenches and non-equilibrium dynamics are penetrating into many fields of physics, including condensed 
matter, cold atoms, high energy physics, mesoscopic systems etc.\cite{dziarmagareview,polkovnikovrmp}. Considering 
the non-equilibrium time evolution not only can one address fundamental questions related to equilibration and 
thermalization, but reaching novel states of matter without an equilibrium counterpart \cite{cayssol,lindner},  
such as Floquet topological systems without an external drive\cite{foster}, also becomes possible.

Strongly correlated systems with short range interaction contain plenty of rich physics, in spite of being 
notoriously difficult to deal with in dimensions higher than one, especially when driven out of equilibrium
after a quantum quench. Mean field models, on the other hand, while 
easily solvable, neglect the interesting physics of  quantum fluctuations.
Models with long range interaction bridge between them and  combine the best of the two by containing interaction terms, and yet being 
exactly solvable in any dimension, at least in the thermodynamic limit.
Such models belong to the family of Richardson-Gaudin  models\cite{dukelskyrmp,ortiz}.

The Richardson-Gaudin models models are characterized by infinitely long, global  range interactions.
Consequently, the Mermin-Wagner theorem is not applicable and ordering can take place independently of the dimensionality of the problem.
These models thus constitute the extreme opposite limit of short range interactions.
It seems also plausible that interacting models with power law interactions \cite{maghrebi} may lead to 
phases and phase transition characteristic to either short range or infinite range interacting models,
which adds to the importance of Richardson-Gaudin type models.
Last but not least, fermionic systems with infinitely long range interactions have recently ceased to 
be of only academic interest, thanks to experiments on cold atomic systems, which have provided a long-sought physical realization \cite{mottl,landig}.

Quantum quenches in Richardson-Gaudin models have been studied by focusing on the fate of a superconducting state upon 
an abrupt change (i.e quantum quench) 
of the interaction parameter\cite{barankov1,barankov2,yuzbashyan,yuzbashyan1,faribault}. The ensuing dynamics 
is expected to be universal  due to the long relaxation time in cold atomic systems, where such experiments were conducted. 
This approach was also extended to the presence of multiple, \emph{fully gapped} order parameters developing on top of  topologically \emph{trivial} non-interacting band structures\cite{moor,fu,dzero}.

Our goal here is to add another twist to the story by considering time evolution in the presence of long range repulsive interactions around a quadratic band touching crossing point, carrying a Berry phase of $2\pi$.
Such systems have been intensively studied in equilibrium and in the presence of short range interactions \cite{kaisun,kunyang,nandkishore,doraherbutmoessner} 
that yield a topologically ordered, quantum anomalous Hall state (QAH)~\cite{hasankane}, which typically wins over the nematic phase.
We find, in contrast, 
that the nematic state competes successfully against the QAH in the presence of long range interaction, and actually becomes more 
robust against quantum quenches than QAH. We attribute the latter feature to the specific topology of the non-interacting dispersion.

\paragraph{Model.}
The model we focus on is the generalization of that in Ref. \cite{kaisun} in the presence of long range interactions.
The kinetic energy in the second quantized form is
\begin{gather}
H_0=\sum_{\bf p}\epsilon_{\bf p}\left(\cos(2\varphi_p)S^x_{\bf p}+\sin(2\varphi_p)S^y_{\bf p}\right),
\label{h0}
\end{gather}
 $\epsilon_{\bf p}=p^2/2m$ and $m>0$ is the effective
mass. 
The resulting non-interacting spectrum describes a quadratic band touching at $p=0$ as $\pm \epsilon_{\bf p}$ and is 
characterized by a non-trivial Berry phase of $2\pi$\cite{novoselov-bilayer-iqhe}
The infinitely long range interactions is
\begin{gather}
H_{int}=-\sum_{i=x,y,z}\frac{g_i}{4N}\left(\sum_{\bf p}S^i_{\bf p}\right)^2.
\label{hint}
\end{gather}
Here, $N$ is the number of unit cells in the system, $S^i_{\bf p}=\Psi_{\bf p}^+\sigma_i\Psi_{\bf p}$ is the analogue of Anderson's pseudospin operator\cite{anderson} in the
particle-hole channel
with $\Psi^+_{\bf p}=(a^+_{\bf p},b^+_{\bf p})$, $\sigma$'s are Pauli matrices, $a^+_{\bf p}$ and $b^+_{\bf p}$ are creation operators of electrons with momentum $\bf p$ on two distinct sublattices\cite{kaisun,doraherbutmoessner,kunyang}.
Due to the long range nature of the interaction, distinct coupling constants are possible for the pseudospin components, in contrast  to the case of short range interactions, which allows only one (marginally) relevant
coupling constant. Each spin interacts with all others in the momentum space, so that the number of nearest neighbor spin components in momentum space is infinity. The self-consistent mean field theory this way becomes exact in the thermodynamic limit.

In this limit, the instabilities manifest themselves as finite expectation values of the spin operators.
In particular, similarly to the case of short range interactions, a finite $\langle S^z\rangle=\sum_{\bf p}\langle S^z_{\bf p}\rangle/N$ order parameter breaks a discrete
symmetry, gaps out the full spectrum and yields the quantum anomalous Hall order. %, and also charge density wave between the two sublattices.
A state with a finite QAH order parameter is topologically non-trivial, and yields
quantized Hall response as $\sigma_{xy}=\pm e^2/h$, irrespective of  the presence or absence of nematic order.
A finite $\langle S^{x,y}\rangle=\sum_{\bf p}\langle S^{x,y}_{\bf p}\rangle/N$ corresponds to nematic order, and breaks the continuous
$U(1)$ rotational symmetry of the spectrum by splitting the quadratic band crossing into two linearly dispersing Dirac cones at finite momentum, each carrying a Berry phase of $\pi$.

\paragraph{Equilibrium phase diagram.}
The physical order parameters appear also in the spectrum through $\Delta=g^z\sum_{\bf p}\langle S^z_{\bf p}\rangle/2N$
and $m_{x,y}=g^{x,y}\sum_{\bf p}\langle S^{x,y}_{\bf p}\rangle/2N$.
For the sake of simplicity, we assume that nematic order develops in the $x$ direction. The order parameters are found by minimizing the total energy of the system
\begin{gather}
E_0=-\rho \int\limits_0^W d\epsilon \int\limits_0^{\pi}\frac{d\phi}{\pi}\tilde\epsilon+\frac{\Delta^2}{g_z}+\frac{m_x^2}{g_x},
\label{e0}
\end{gather}
with $\tilde\epsilon=\sqrt{\epsilon^2+\Delta^2+m_x^2-2\epsilon m_x\cos(\phi)}$,  $\rho=m/2\pi$ is the density of states, $W$ is the high energy cutoff. For $g_x=g_y$, the angle of the order
parameter remains undetermined from the mean-field equations, and becomes the Goldstone mode of the continuous rotational symmetry breaking.
For $g_x\gtrless g_y$, the bare interaction itself breaks the rotational symmetry and nematic order, breaking now a discrete $Z_2$ symmetry, develops only in the $x/y$ direction, respectively. 

\begin{figure}[h!]
\psfrag{x}[t][][1][0]{$\rho g_z$}
\psfrag{y}[b][][1][0]{$\rho g_x$}
\includegraphics[width=4.4cm]{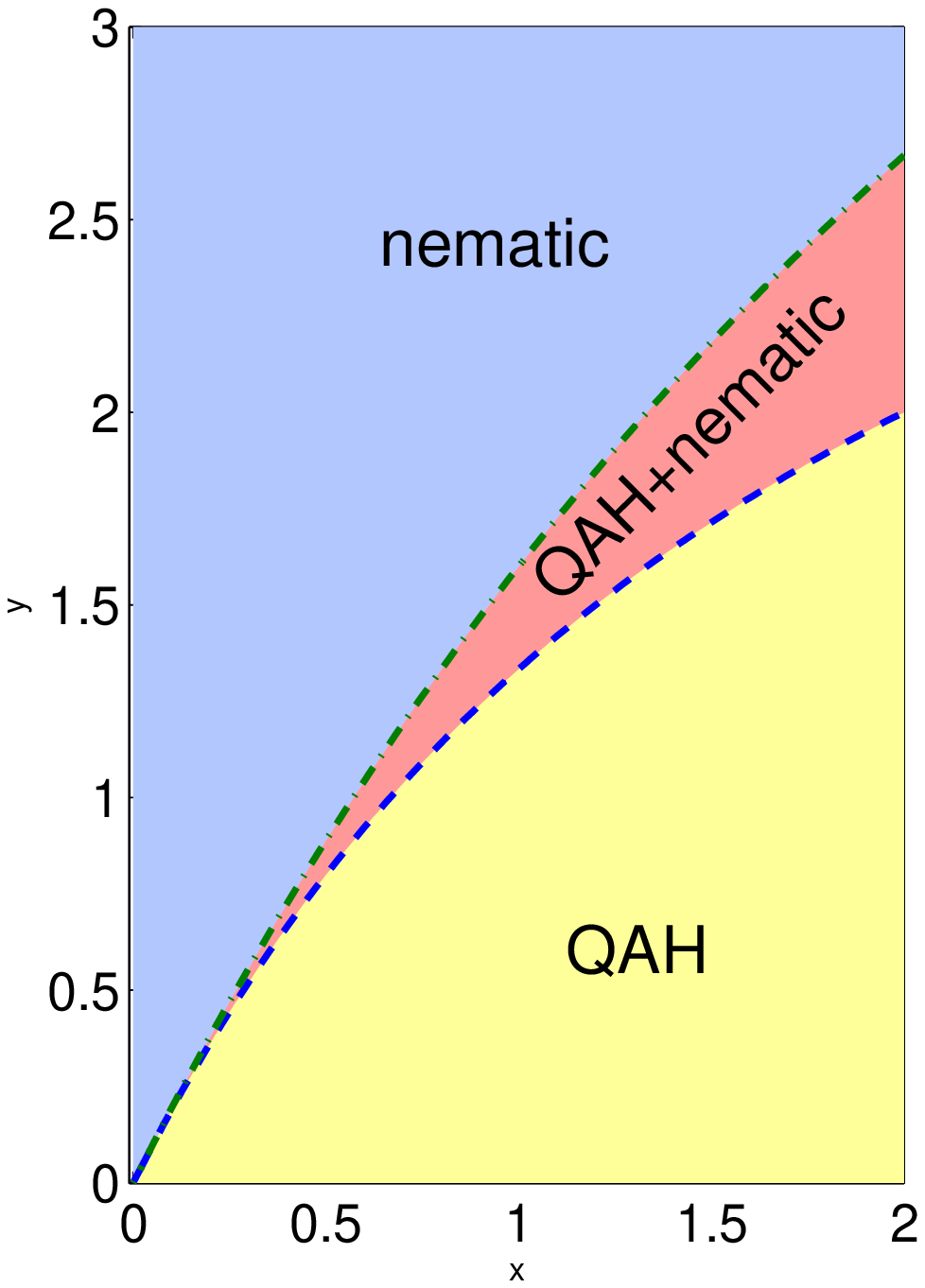}
\psfrag{x}[t][][1][0]{$\rho g_x$}
\psfrag{y}[b][][1][0]{\color{blue} $\Delta$, \color{red} $m_x$}
\includegraphics[width=4cm]{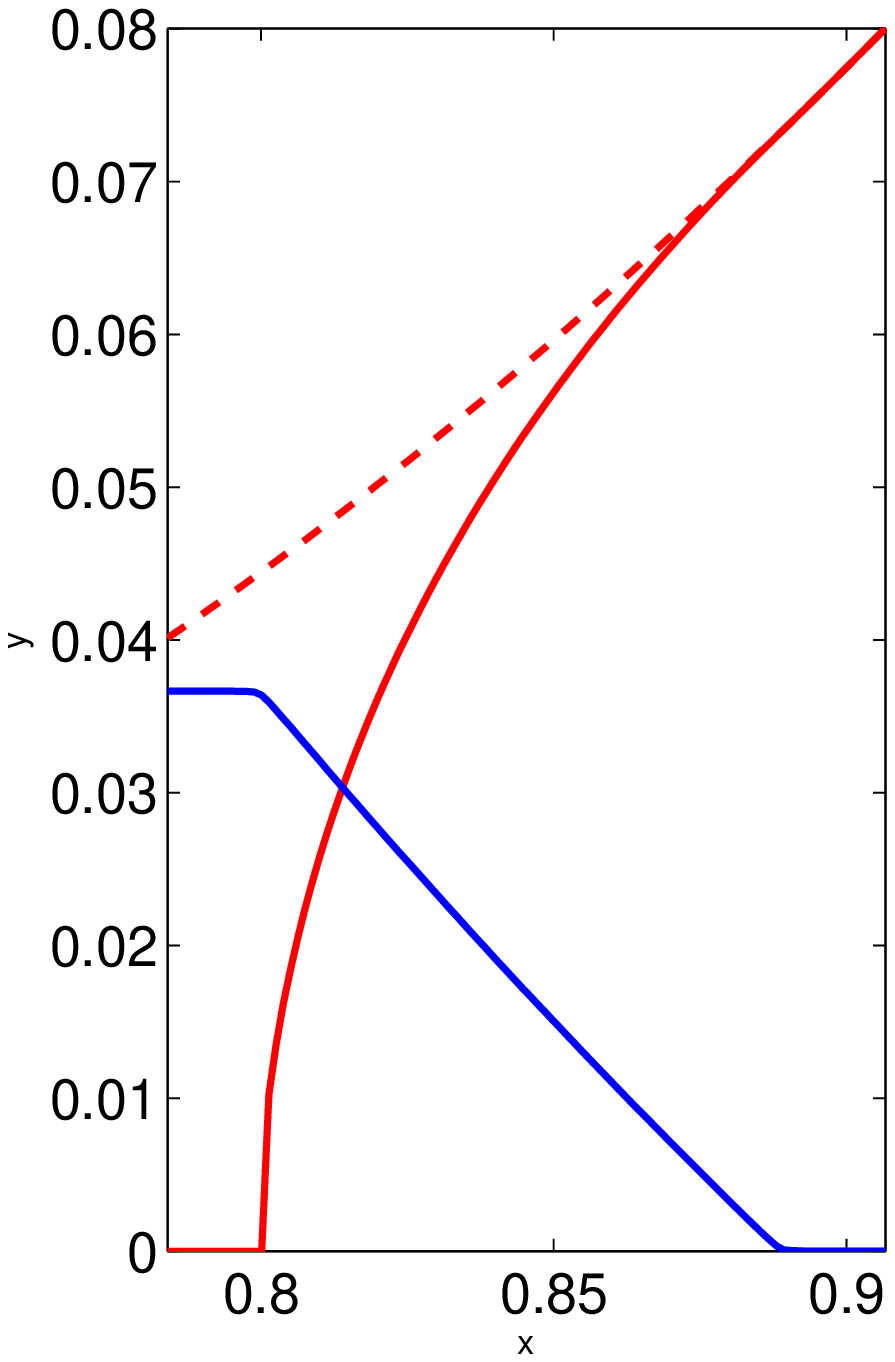}
\caption{(Color online) The weak coupling phase diagram of Eqs. \eqref{h0}+\eqref{hint} for $g_x\geq g_y$ in the weak-coupling limit is shown in the left panel.
For $g_y>g_x$, $g_x$ should be replaced by $g_y$. The blue dashed/green dash-dotted lines denote a second/topological third order transition, respectively.
Right panel: the nematic (red) and QAH (blue) order parameters are shown for $\rho g_z=1/2$ in the coexisting region. The red dashed line denotes the nematic gap without QAH,
which is only a local minimum of the ground state energy compared to the coexisting solution.}
\label{phasediag}
\end{figure}

Minimizing the energy in Eq. \eqref{e0} yields the gap equations as
\begin{gather}
(\Delta,m_x)=\rho \int\limits_0^W d\epsilon \int\limits_0^{\pi}\frac{d\phi}{\pi}\frac{(g_z\Delta,g_x(m_x-\epsilon\cos(\phi))}{2\tilde\epsilon}.
\label{mfeqs}
\end{gather}
The phase diagram emerging from their solution is shown in Fig. \ref{phasediag}, which is our first important result.
For $g_z\gg g_x$, the QAH phase suppresses nematic order, while in the opposite, $g_x\gg g_z$ situation, the nematic order wins. In between, two
phase boundaries are identified, corresponding to continuous phase transitions.
 The $g_x=4g_z/(2+\rho g_z)$ line in the weak coupling limit marks a second order transition according to Ehrenfest classification and separates the pure QAH state from a coexisting QAH and nematic phase. The nematic order parameter rises as $m_x\sim \sqrt{|g-g_c|}$, and the ground state energy changes as $E_0\sim m_x^4$, so that its second derivative is discontinuous.

The $g_x=8g_z/(4+\rho g_z)$ weak coupling line, on the other hand, denotes
a topological, third order transition from the phase with coexisting orders to a pure nematic state.
The QAH order varies as $\Delta\sim |g-g_c|$, producing $E_0\sim \Delta^3$ with a jump in the third derivative.
The third order transition is characteristic
to the mean field Dirac metal-insulator transition, as noted already in Ref. \cite{sorella}.
A third order topological transitions occurs
also in related long range interacting model\cite{ortiz}.
It is an interesting feature of our model that not only nematic order appears in the weak coupling limit,
but it also coexists with QAH for vanishingly small couplings, in sharp contrast to the case of short range interaction\cite{kaisun}.
In the non-coexisting regions, the conventional weak coupling forms are recovered as $\Delta=2W\exp(-2/\rho g_z$) and $m_x=4W\exp(\frac 12-4/\rho g_x)$.

\paragraph{Quantum quench.}
Having determined the equilibrium
properties of $H_0+H_{int}$, we turn to its behaviour after an interaction quantum quench. The model initially sits in the ground state for some coupling parameters, which changes abruptly to some other value at $t=0$.
The ensuing dynamics is governed by the time-dependent mean field theory\cite{barankov1,barankov2,yuzbashyan,yuzbashyan1}, yielding
\begin{subequations}
\begin{gather}
\partial_t S^x_{\bf p}=2\epsilon_{\bf p}\sin(2\varphi_p)S^z_{\bf p}+g_zS^z(t) S^y_{\bf p}-g_y S^y(t) S^z_{\bf p},\\
\partial_t S^y_{\bf p}=-2\epsilon_{\bf p}\cos(2\varphi_p)S^z_{\bf p}+g_x S^x(t) S^z_{\bf p}-g_z S^z(t) S^x_{\bf p},\\
\partial_t S^z_{\bf p}=2\epsilon_{\bf p}\left(\cos(2\varphi_p)S^y_{\bf p}-\sin(2\varphi_p)S^x_{\bf p}\right)+\nonumber\\
+g_yS^y(t)S^x_{\bf p}-g_xS^x(t) S^y_{\bf p},
\end{gather}
\label{meanfieldt}
\end{subequations}
where $S^i(t)=\sum_{\bf p}\langle S^i_{\bf p}\rangle /N$ is the time evolved order parameter after the quench.
The initial conditions at $t=0$ are
\begin{gather}
\left(\begin{array}{c}
\langle S^x_{\bf p}\rangle \\
\langle S^y_{\bf p}\rangle  \\
\langle S^z_{\bf p}\rangle
\end{array}\right)
=
\left(\begin{array}{c}
m_x-\epsilon_{\bf p}\cos(2\varphi_p)\\
-\epsilon_{\bf p}\sin(2\varphi_p)\\
\Delta
\end{array}\right)\times E_{\bf p}^{-1},
%\langle S^x_{\bf p}\rangle=\frac{m_x-\epsilon_{\bf p}\cos(2\varphi_p)}{\sqrt{\epsilon_{\bf p}^2+\Delta^2+m_x^2-2\epsilon_{\bf p} m_x\cos(2\varphi_p)}},\\
%\langle S^y_{\bf p}\rangle=-\frac{\epsilon_{\bf p}\sin(2\varphi_p)}{\sqrt{\epsilon_{\bf p}^2+\Delta^2+m_x^2-2\epsilon_{\bf p} m_x\cos(2\varphi_p)}},\\
%\langle S^z_{\bf p}\rangle=\frac{\Delta}{\sqrt{\epsilon_{\bf p}^2+\Delta^2+m_x^2-2\epsilon_{\bf p} m_x\cos(2\varphi_p)}}
\end{gather}
where $E_{\bf p}=\sqrt{\epsilon_{\bf p}^2+\Delta^2+m_x^2-2\epsilon_{\bf p} m_x\cos(2\varphi_p)}$.

Eqs. \eqref{meanfieldt} are solved numerically using e.g. a fourth order Runge-Kutta method.
Since the full parameter space is large due to the distinct coupling constants in Eq. \eqref{hint}
before and after quench, we focus on the time evolution of the pure QAH and nematic order parameters,
when only their respective coupling constants are present and quenched.

\begin{figure}[h!]
\psfrag{x}[t][][1][0]{$tW$}
\psfrag{y}[b][][1][0]{\color{blue} $S^z(t)$ (QAH), \color{red} $S^x(t)$ (nematic)}
\includegraphics[width=7.5cm]{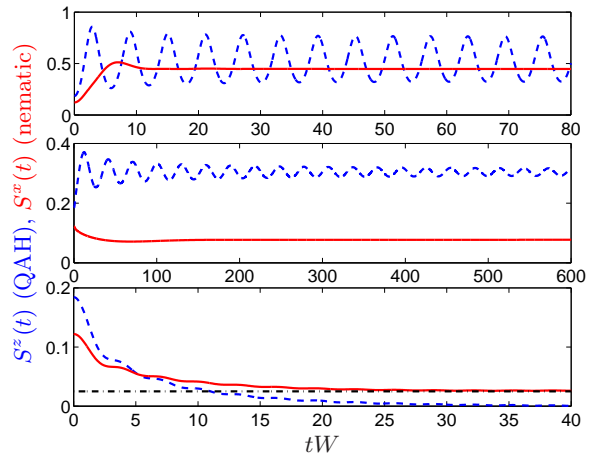}
\caption{(Color online) The time evolution of the QAH (blue dashed) and nematic (red solid) order parameters are shown, starting from $\Delta=0.05W$ or $m_x=0.05W$,
corresponding to an initial coupling $\rho g_z=0.543$ or
$\rho g_x=0.819$, respectively. After the quench,  $\rho g_{z,x}=2$ (top panel),
0.7 (middle panel) and 0.1 (bottom panel).
For the latter, the steady state value is indicated by a black dash-dotted from Eq. \eqref{nematicss}.
The tiny oscillation superimposed on the temporal
decay arises from a finite cutoff, which is needed to make $S^{x,z}(t=0)$ finite. }
\label{opvst}
\end{figure}

\paragraph{QAH quench.}
In the pure QAH case with $g_{x,y}=0$, the phase variable $\varphi_p$ can be transformed away by a rotation around $z$ for each spin, leaving us with an effective one dimensional model of the Richardson-Gaudin
type\cite{dukelskyrmp}, which depends only on the momentum $p$, and which is exactly solvable.
The interaction, however, here is  of 'easy axis' type, e.g. $-(S^z)^2$, in contrast to the usual 'easy plane' interaction in the BCS theory, which is  $-(S^x)^2-(S^y)^2$.
In other words, the QAH order parameter is always real and associated with breaking of a discrete symmetry, whereas the BCS order parameter breaks the continuous $U(1)$ symmetry. Nevertheless, the emerging picture resembles closely that of a quenched BCS superconductor, and reveals three qualitatively distinct regions in the temporal dynamics, as shown in Fig. \ref{opvst}. For large final coupling, persistent, non-decaying
oscillations show up in the time dependence, and this phase-locked, self-induced, synchronized oscillation can be used to realize an externally non-driven Floquet topological phase, similarly to p-wave superconductors\cite{foster}.
For medium values of the final coupling, the QAH order parameter reaches a time independent steady state value through power-law decaying ($\sim t^{-1/2}$) 
damped oscillations, while for small quenches, an exponential decay to zero occurs.
Note that tiny additional oscillations of the form $\sin(2Wt)/Wt$ are superimposed on the decay due to the finite cutoff $W$, which is essential to keep the equilibrium $S^z(t=0)$ finite. These three regions are separated by dynamical phase transitions\cite{sciolla,barankov2}.

\paragraph{Nematic quench.}
The quench of a pure nematic state with $g_{y,z}=0$, on the other hand, is characterized by completely different behavior. 
No dynamical phase transition occurs at all, and the time evolution is characterized by damped oscillations with a much faster decay when compared to the QAH case. This is
 due to the existence of low energy excitations around the linearly dispersing Dirac points, as visualized in Fig. \ref{opvst}.
The crucial difference with respect to previously considered quenches in superconductors \cite{barankov2,peronaci} occurs, however, for small finite couplings.
In this limit, the nematic order parameter does not vanish but approaches a finite, time independent
stationary value. This is best illustrated by completely switching-off the nematic coupling, in which case Eqs. \eqref{meanfieldt} admit a simple solution as
\begin{gather}
\langle S^x_{\bf p}\rangle=\frac{m_x[\cos(2\epsilon_{\bf p}t)\sin^2(2\varphi_p)+\cos^2(2\varphi_p)]-\epsilon_{\bf p}\cos(2\varphi_p)}{\sqrt{\epsilon_{\bf p}^2+m_x^2-2\epsilon_{\bf p} m_x\cos(2\varphi_p)}},
\end{gather}
yielding, after momentum integration, in the stationary state when  $t\rightarrow \infty$
\begin{gather}
S^x_{st}=\frac{\rho m_x}{2},
\label{nematicss}
\end{gather}
with the superimposed power-law decay, $\sim \rho \cos(2m_xt)/m_x^2t^3$, obtained through the method of steepest descent. The exponent of the power-law differs both from those in s-wave\cite{gurarie} or d-wave\cite{peronaci} superconductors, in spite of the  latter also possessing low-energy Dirac-like quasiparticles.
This reflects both the linearly dispersing Dirac cones and the nontrivial $2\pi$ Berry phase and results in heavily damped oscillations in Fig. \ref{opvst}.
Since no qualitative change occurs in the time evolution of the order parameter with increasing final coupling, this supports the idea\cite{yuzbashyan1} that both the final stationary value together with the $\sim t^{-3}$ decay is a universal feature of the nematic phase.

Eq. \eqref{nematicss} is our second main result. It is surprising for several reasons: first, the nematic order is always suppressed
for short range interactions in equilibrium\cite{kaisun}, and although it competes with QAH successfully
for long range interactions, it would be naively expected to be more vulnerable to time dependence of the interaction coupling,  due to
its low energy excitations. Second, the equilibrium low-energy thermodynamics of 2D Dirac fermions is universal, no matter
whether it is for graphene\cite{castro07}, d-wave superconductors\cite{wonmaki} or the present nematic state.
In sharp contrast, the dynamics after quench differs significantly from that of a d-wave BCS superconductor\cite{peronaci} where the
order parameter vanishes identically for small quenches.
 We trace this difference back to the non-trivial non-interacting spectrum and its $2\pi$ Berry phase in Eq. \eqref{h0}, as well
as to the structure of the interaction, both of which conspire to produce the nematic order.
In contrast, d-wave superconductivity is induced only by the symmetry of the interaction\cite{peronaci} and is largely insensitive to the bare dispersion; 
its stationary state therefore resembles closely that of an s-wave superconductor.

Third, the survival of the nematic order does \emph{not} simply 
follow from energetics\cite{barankov2}, namely, from comparing the post-quench energy of the system to
that of an equilibrium nematic state, to estimate its effective temperature.
For an interaction switch-off in the weak coupling limit, the energy considerations are
identical to that in a d-wave superconductor, where, in contrast to Eq. \eqref{nematicss},
 the order parameter vanishes\cite{peronaci}. In spite of the deduced effective temperature being above 
 the equilibrium critical temperature for the nematic ordering, nematicity survives the quench.
The essential difference between nematic and superconducting states lies in the pseudospin structure.
An initial spin polarized state in the $x-y$ plane is driven by a pseudomagnetic field in the $z$ direction for a superconductor
or by an in-plane hedgehog like field configuration from $H_0$ for the nematic state.
While the former produces complete dephasing, the latter dephases only partially,  and still preserves some information of the initial state.

 We also find that starting from an initial coexisting nematic and
QAH phase and switching off the interactions, the nematic order survives according to Eq. \eqref{nematicss} while the QAH vanishes.

%\begin{figure}[h!]
%\psfrag{x}[t][][1][0]{$tW$}
%\psfrag{y}[b][][1][0]{\color{blue} $S^z(t)$, \color{red} $S^x(t)$, \color{black}  $S^y(t)$}
%\psfrag{t1}[][][0.8][0]{$\rho g_x=0.94$, $\rho g_y=0$, $\rho g_z=0.6$}
%\psfrag{t2}[][][0.8][0]{$\rho g_x=\rho g_y=0.94$, $\rho g_z=0.6$}
%\includegraphics[width=8.6cm]{coexistquench.eps}
%\caption{(Color online) Quantum quenches within the coexisting region, starting from $m_x=\Delta=0.03 W$ and with easy axis nematic (left panel) or easy plane nematic
%(right panel) interactions.}
%\label{coexistquench}
%\end{figure}

We have so far focused on 'easy axis' nematic quenches with $g_x\neq 0$ and $g_y=g_z=0$, but
our results apply equally to 'easy plane' nematic quenches with $g_x>g_y>0$ and $g_z=0$, when
only a discrete symmetry is broken in equilibrium by the nematic order and the ground state energy possesses a discrete number of minima.
On the other hand, for $g_x=g_y$, a continuous symmetry is broken, the equilibrium ground state energy has a Mexican hat structure, with
all of the minima ending up being mixed after a quench. Consequently, the above solution with only $S^x\neq 0$ is unstable with respect to
any small $S^{y,z}$ additional order parameters.

%Finally, let us remark on the coexisting phase. The spectrum is gapped due to the coupled nematic and QAH order parameters in
%Eq. \eqref{meanfieldt},
%and all order parameters exhibit oscillations around their steady state values after a quench, similar to other gapped systems\cite{moor,dzero}.
%Representative  numerical results are shown in Fig. \ref{coexistquench}.

\begin{figure}[h!]
\psfrag{x}[t][][1][0]{$\rho g_{x,z}$}
\psfrag{y}[b][][1][0]{\color{blue} $S^z_{st}$ (QAH), \color{red} $S^x_{st}$ (nematic)}
\includegraphics[width=6cm]{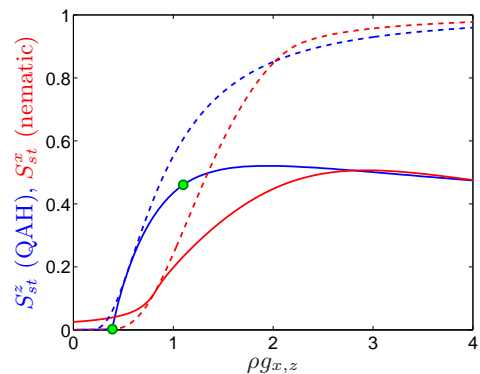}
\caption{(Color online) The steady state behaviour of the order parameter is plotted for the pure QAH (blue) and nematic (red) state, starting from either
 $\Delta=0.05W$ or $m_x=0.05W$ in the respective channel. The green dots denote the three qualitatively different regions of the QAH order parameter.
The dashes lines are the equilibrium order parameters from the solution of Eqs. \eqref{mfeqs}.}
\label{gapvsg}
\end{figure}

\paragraph{Steady state.}
Based on these, we can construct the stationary state phase diagram of the system, shown in Fig. \ref{gapvsg}. The stationary values of the nematic and QAH order behave qualitatively similarly, expect for small quenches where
only the nematic order survives. The large coupling region of the QAH phase is also accompanied by persistent oscillations, which are absent in the nematic case.

\paragraph{Discussion.}
We have studied the interplay of nematic and QAH orders around a 2D quadratic band touching point in the presence of long range interactions.
In equilibrium, nematic order occupies a significant portion of the phase diagram, and can even coexist with QAH through a third order quantum phase transition, before
giving way to QAH order via a second order phase transition.
After a quantum quench, the gapped QAH order behaves similarly to a BCS superconductor, and vanishes in the steady state for small final couplings.
Surprisingly, the gapless  nematic order survives any quenches and remains finite in the steady state, due to the peculiar topology of the quadratic band touching, and defying the usual simple energetic reasoning.

\begin{acknowledgments}

BD is supported by the Hungarian Scientific Research Fund
Nos. K101244, K105149, K108676 and by the Bolyai Program of the HAS. IFH is supported by the NSERC of Canada.
\end{acknowledgments}

\bibliographystyle{apsrev}
\bibliography{refgraph}

\end{document}